\documentclass[twocolumn,pr,notitlepage,longbibliography,superscriptaddress]{revtex4-2}
\usepackage{mathtools}
\usepackage{graphicx}
\usepackage{mathrsfs}
\usepackage{amssymb}
\usepackage{soul}
\usepackage{xcolor}
\usepackage[colorlinks=true,citecolor=blue,urlcolor=blue]{hyperref}
\usepackage{amsmath}
\usepackage{bm}
\usepackage{amsmath}
\usepackage{physics}

\usepackage{amsmath}

\newcommand{\mysection}[1]{\textcolor{blue}{\textit{#1}.}}

\begin{document}

\title{Spin orientation by electric current in altermagnets}

\author{L. E. Golub}
\affiliation{Physics Department, University of Regensburg, 93040 Regensburg, Germany}
\author{L. \v{S}mejkal}
 \affiliation{Max Planck Institute for the Physics of Complex Systems, N\"othnitzer Str. 38, 01187 Dresden, Germany}
\affiliation{Max Planck Institute for Chemical Physics of Solids, N\"othnitzer Str. 40, 01187 Dresden, Germany} 
\affiliation{Institute of Physics, Czech Academy of Sciences, Cukrovarnická 10, 162 00 Praha 6, Czech Republic}

\begin{abstract}
{We theoretically demonstrate that the flow of electric current in an altermagnet results in the formation of a homogeneous electron spin orientation. 
We show that the spin of the conduction electrons generated in altermagnets with $d$-wave spin-momentum couplings,
is quadratic in the current magnitude, varies as the second angular harmonic under variation of the current direction and does not require broken inversion symmetry.
The effect is thus distinct from conventional current induced spin polarization phenomena which are linear in the current, vary as a first angular harmonic under variation of current direction and require broken inversion symmetry.
We analytically derive the current-induced spin orientation in altermagnets using the kinetic theory for distribution functions in the spin-split subbands. Finally, we show numerically that
an application of external magnetic field significantly enhances the induced electron spin.
}
\end{abstract}

\maketitle

\mysection{Introduction}
Altermagnetism is arguably one of the most rapidly growing parts of condensed matter physics. After a recent understanding that the altermagnetic spin symmetry class is different from conventional ferro- and antiferromagnetic ones, a big avalanche of research appeared focusing on electronic phenomena in altermagnets~\cite{Smejkal2022,Smejkal2022a,Mazin2022,Bai2024,Krempasky2024,Reimers2024,Amin2024} 
and a large palette of altermagnetic materials was proposed~\cite{Smejkal2022,Smejkal2022a,Bai2024}. The material candidates include both three-dimensional systems, e.g. MnTe and CrSb~\cite{Smejkal2022,Feng2022}, and two-dimensional monolayers, e.g. RuF$_4$ and Fe(S,Se)~\cite{Smejkal2022a,Mazin2023,Milivojevic2024}. The MnTe and CrSb altermagnets were already experimentally confirmed in photoemission~\cite{Krempasky2024,Reimers2024} and x-ray magnetic circular dichroism~\cite{Amin2024} studies.

The spontaneous altermagnetic ordering is described by a spin group which breaks spin-space and lattice-space rotation symmetries but preserves their certain combinations~\cite{Smejkal2022}. Consequently, the direct space magnetization densities and momentum space spin order of altermagnets exhibit $d$-, $g$-, or $i$-wave symmetries which makes them distinct symmetry class from conventional ferromagnets and antiferromagnets\cite{Smejkal2022}. The altermagnetic compensated spin order in momentum space also breaks time-reversal symmetry~\cite{Smejkal2020} and leads to many intriguing physical properties which cannot be found in conventional antiferromagnets such as unconventional anomalous Hall effect~\cite{Smejkal2020} and giant magnetoresistance~\cite{Smejkal2022b} without magnetization, and multiferroic and altermagnetoelectric couplings~\cite{Smejkal2022a,Gu2024,Smejkal2024} with potential applications in nanoelectronics, spintronics and multiferroics~\cite{Smejkal2022a,Bai2024}.

\begin{figure}[h]
	\centering 
	\includegraphics[width=0.42\linewidth]{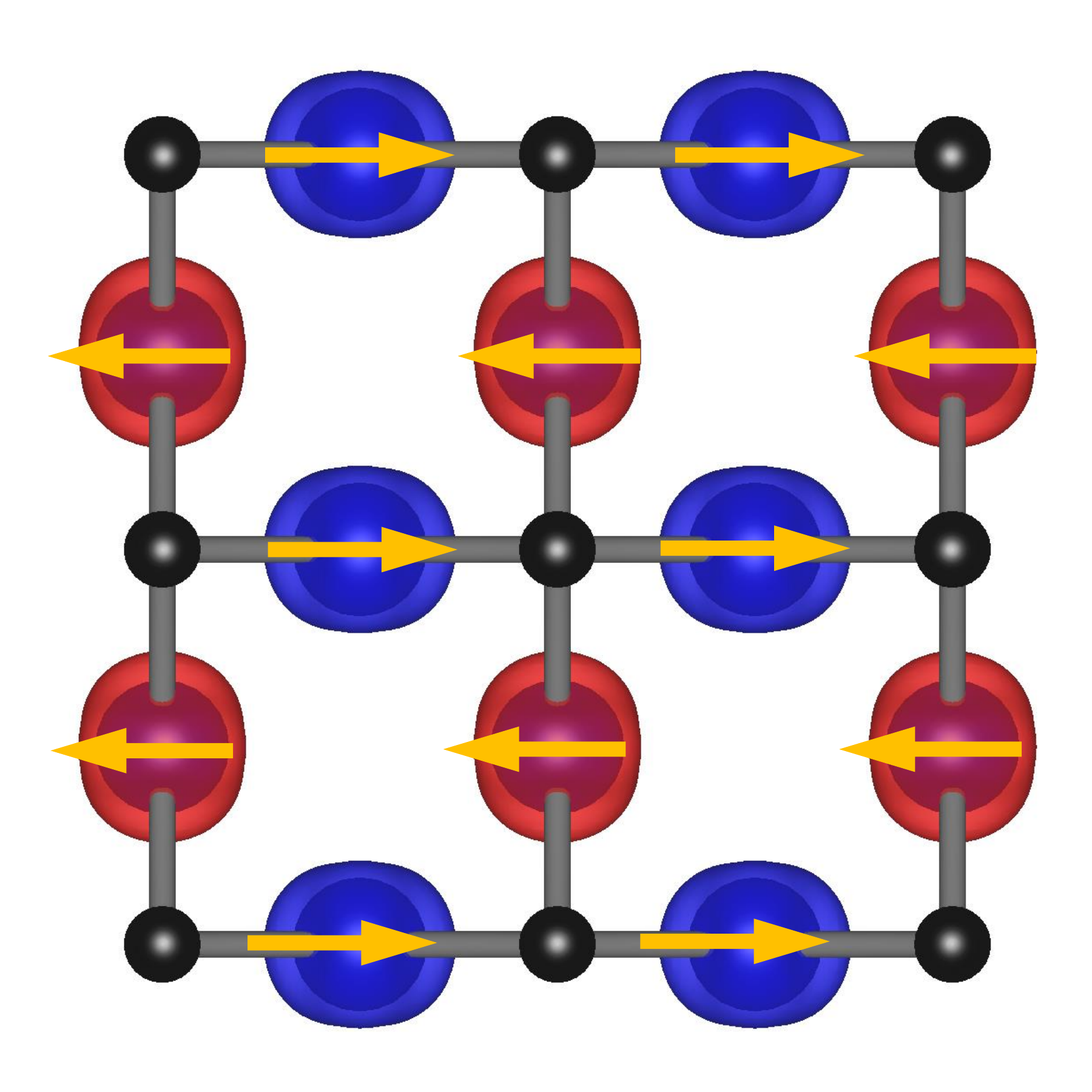}
	\includegraphics[width=0.5\linewidth]{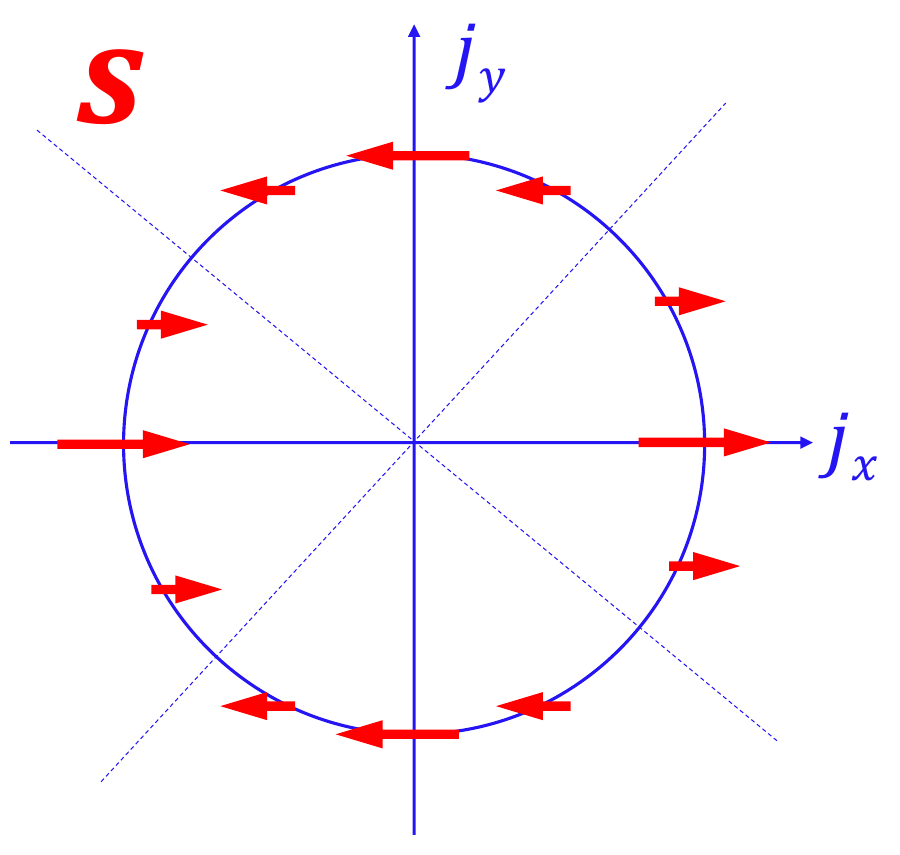} 
	\caption{ 
	Left: Alternating magnetic and crystal pattern in a $d$-wave altermagnet. Right: Directions of electron spin oriented by electric current.
	The electron spin is shown by red arrows. For $\bm j \parallel x$ the net spin polarization $s_x>0$, while at $\bm j \parallel y$ we have  $s_x<0$, see Eq.~\eqref{s_j_phenom}.
	}
	\label{fig0}
\end{figure}

In the present letter, we theoretically demonstrate that the symmetry of altermagnets with quadratic-in-momentum spin-momentum locking allows for the following  phenomenon: electrical current passing through a sample results in a net electron spin polarization.
In contrast to the spin-Hall effect, where the spin is oppositely oriented at the sample edges~\cite{Bai2023}, the electrical spin orientation 
(called in literature also current induced spin polarization, Edelstein or inverse-spin galvanic effect) results in the homogeneous spin orientation in the whole sample.
In semiconductors and semiconductor heterostructures, the  induced spin orientation is linear in the electric current, for reviews see Refs.~\cite{Ganichev2018,Ganichev2024a} and references therein. 
Analogical effects, with induced spin orientation linear in the electric current, were proposed in
certain non-collinear magnets~\cite{GonzalezHernandez2024,Hu2024,Chakraborty2024}.
By contrast, the induced electron spin in $d$-wave altermagnets is quadratic in the electric current. The following phenomenological relation is allowed by symmetry between components of the electric current density $\bm j$ and the electron spin orientation degree $\bm s$
\begin{equation}
\label{s_j_phenom}
s_x = Q (j_x^2 - j_y^2),
\end{equation}
where $x$,~$y$ are the main axes of $d$-wave symmetry, and $x$ is a spin orientation direction in the 
model altermagnet lattice,
Fig.~\ref{fig0}, inspired by realistic monolayer materials~\cite{Smejkal2022,Mazin2023,Fernandes2024,Parthenios2025}.

We use a two-dimensional model~\cite{Smejkal2022,Mazin2023,Fernandes2024,Denisov2024} and take the Hamiltonian of an altermagnet in the following form
\begin{equation}
\label{H}
\mathcal H = {\hbar^2 k^2\over 2m} + \beta (k_x^2-k_y^2)\sigma_x.
\end{equation}
Here $\bm k$ is quasimomentum, $m$ is the electron effective mass, $\sigma_x$ is the spin Pauli matrix,
and $\beta$ is the parameter responsible for the altermagnetism.

In the altermagnets with the Hamiltonian~\eqref{H}, 
we can estimate the coefficient $Q$ in Eq.~\eqref{s_j_phenom}  from an approximate relation ${s_x \approx \beta (k_{{\rm dr},x}^2-k_{{\rm dr},y}^2)/\varepsilon_{\rm F}}$, where $\varepsilon_{\rm F}$ is the Fermi energy, and $\bm k_{\rm dr}$ is the drift wavevector related with the electric current  and the electron concentration $N$ as $eN \hbar \bm k_{\rm dr}/m = \bm j$. This yields 
\begin{equation}
\label{Q_estimate}
Q \approx {\beta \over \varepsilon_{\rm F}} \qty({m\over e\hbar N})^2.
\end{equation}
Hereafter we assume that the spin splitting $\sim 
\beta k^2$ is much smaller than the kinetic energy, i.e. that ${m\beta/\hbar^2 \ll 1}$.

In semiconductor quantum wells, a similar effect was proposed~\cite{Tarasenko2006} with a high-frequency electric current orienting the electron spin, but it was caused by the $\bm k$-linear spin-orbit interaction and appears in second order in spin-orbit coupling strength. In contrast, the electrically induced spin in the altermagnets is linear in the constant $\beta$.
Recently, the electron spin orientation by a square of electric field 
has been studied in nonmagnetic, ferromagnetic and antiferromagnetic systems and termed as nonlinear Edelstein effect~\cite{Xiao2022,Xiao2023,Guo2024,Baek2024}. Various contributions to the spin orientation have been identified including intrinsic, extrinsic as well as odd and even in time-reversal. In the present work, we 
calculate the leading, time-reversal even contribution  in altermagnets.

\begin{figure}[t]
	\centering \includegraphics[width=\linewidth]{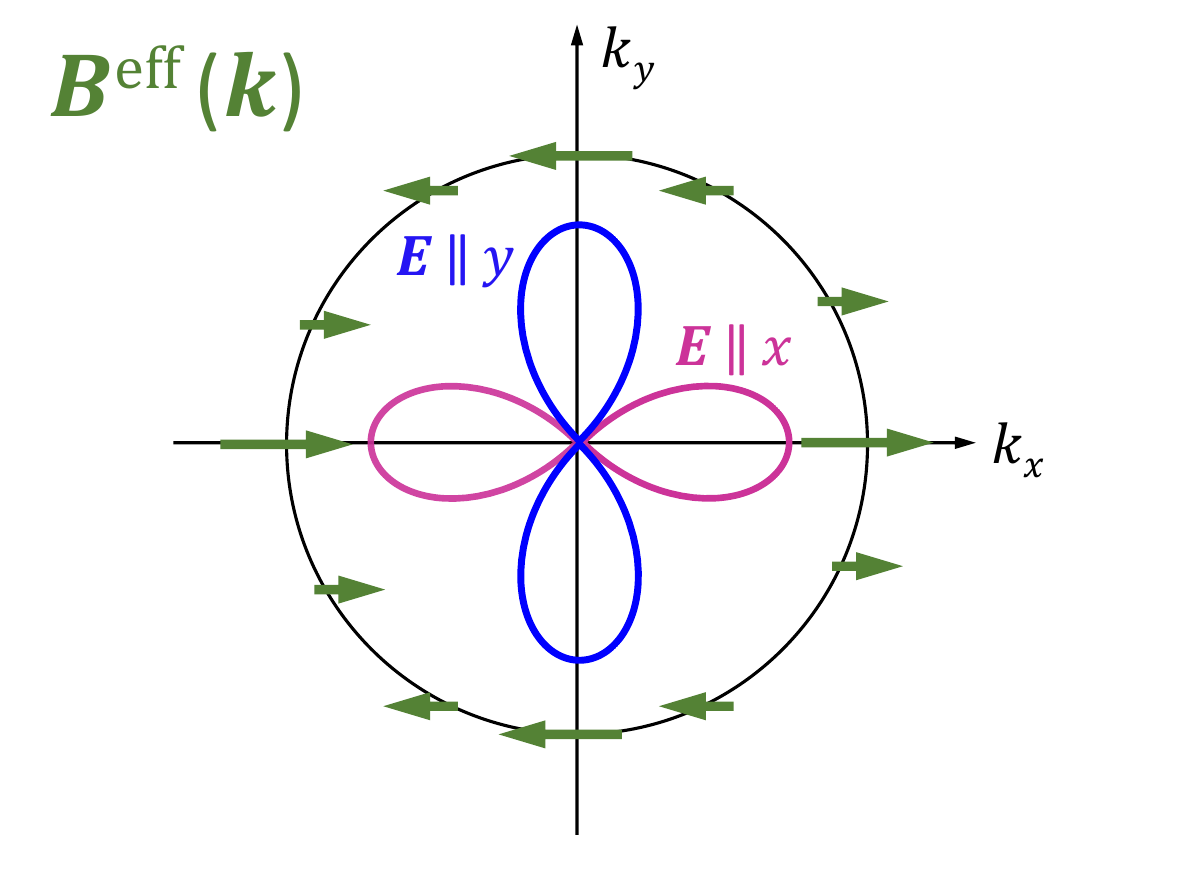} 
	\caption{ 
	Qualitative picture of electron spin orientation in altermagnets. The distribution function in the momentum space has a correction $\delta \tilde{f}_{\bm k}$ quadratic in the electric field strength $\bm E$ with the anisotropy controlled by the direction of $\bm E$ (momentum alignment), Eq.~\eqref{align}. The angular dependence of $\delta \tilde{f}_{\bm k}$ for two orientations of the electric field  is shown by the  magenta and blue curves. The  direction of the effective magnetic field $\bm B^{\rm eff}(\bm k)$ is shown by red arrows. 
	The effective magnetic field in altermagnets has opposite sign of its $x$-component 
for $\bm k \parallel x$ and $\bm k \parallel y$.	
	The momentum alignment results in a higher (lower) occupation of the electron states with horizontally oriented momenta at $\bm E \parallel x$ ($\bm E \parallel y$), and they feel the effective magnetic field  $B^{\rm eff}_x>0$ ($<0$). This results in the net spin polarization $\bm s \parallel x$.
	}
	\label{fig1}
\end{figure}

Qualitatively, the electric spin orientation appears due to the alignment of the electron momenta in the electric field~\cite{Olbrich2014,Otteneder2020}. This is an effect of the anisotropic momentum distribution that appears in the second order in the electric field strength. The distribution function has a correction 
\begin{equation}
\label{align}
\delta \tilde{f}_{\bm k} \propto \abs{\bm E}^2\cos(2\varphi_{\bm k}-2\varphi_{\bm E}), 
\end{equation}
where $\varphi_{\bm k}$ and $\varphi_{\bm E}$ are the polar angles of $\bm k$ and the electric field vector $\bm E$, see Fig.~\ref{fig1}. As a result of this redistribution of the electrons in the momentum space, the number of carriers moving in the direction of $\bm E$ is larger than that for $\bm k \perp \bm E$. 
The second ingredient necessary for the electrical spin orientation is an effective magnetic field $\bm B^{\rm eff}(\bm k)$ present in altermagnets. It arises due to the spin-dependent term in the Hamiltonian~\eqref{H} which can be rewritten as $\bm \sigma\cdot \bm B^{\rm eff}(\bm k)$ with 
\begin{equation}
\label{B_eff}
B^{\rm eff}_x(\bm k)=\beta (k_x^2-k_y^2).
\end{equation}
If $\bm E$ is oriented along $x$ or $y$ axis, the higher number of electrons feel the effective magnetic field $B^{\rm eff}_x$, see Fig.~\ref{fig1}. This results in a net spin polarization $\bm s \parallel x$. By contrast, if $\bm E$ is directed at $\pm 45^\circ$ to the axes $x, y$, then the actual $\bm B^{\rm eff}=0$, and the spin polarization is absent.

Below we calculate the spin orientation degree in a 2D $d$-wave altermagnet.
We assume that the spin energy branches are well-resolved, $\beta k_{\rm F}^2 \gg \hbar/\tau$, where $k_{\rm F}$ is the Fermi wavevector, and $\tau$ is the elastic relaxation time. 

\mysection{Microscopic theory}
Microscopically, the Hamiltonian~\eqref{H} describes two decoupled spin subsystems, and a coupling between them is necessary to obtain a nonzero spin. In order to make this coupling, 
we include an additional term to the Hamiltonian
\begin{equation}
\delta \mathcal H =2\alpha k_x k_y \sigma_y,
\label{dH}
\end{equation}
where $\sigma_y$ is a Pauli matrix different from that in Eq.~\eqref{H}, and $\alpha$ is some parameter. 
This term allows for spin relaxation in the system and results in the nonzero spin orientation degree $s_x$ which is, however, finite at  $\alpha \to 0$.
{The term can arise in certain system from a combined effect of relativistic spin-orbit coupling and nonrelativistic altermagnetism~\cite{Krempasky2024,Fernandes2024,Belashchenko2025}.}

We also allow for an  external magnetization. The Zeeman effect is described by the Hamiltonian
\begin{equation}
\mathcal H_{\rm Z} = \Delta \sigma_z,
\end{equation}
with the Zeeman splitting being equal to $2\Delta$.

\begin{figure}[t]
	\centering \includegraphics[width=\linewidth]{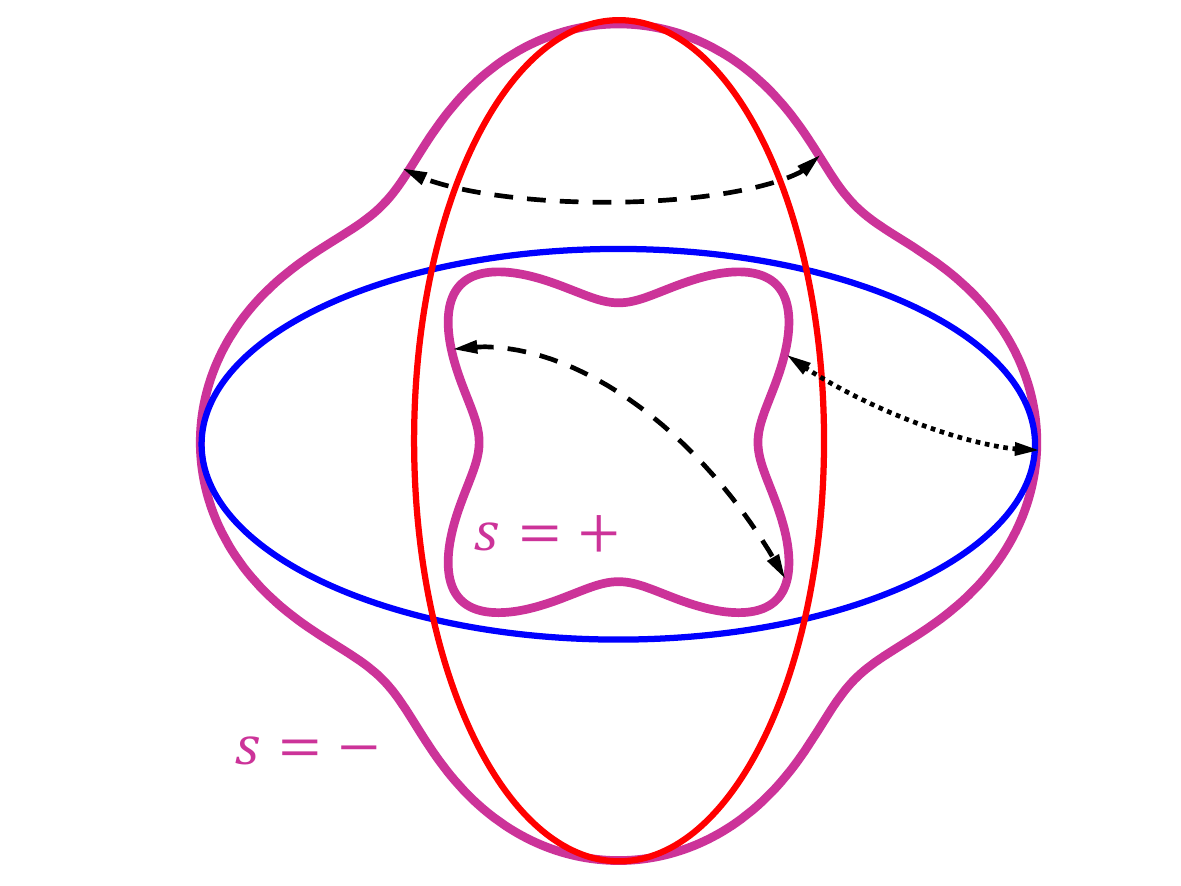} 
	\caption{ 
	Fermi contours in the absence of magnetization ($\Delta=0$) at $\alpha=0$ (blue and red  curves) and at $\beta>\alpha>0$ (solid curves). The microscopic theory accounts for elastic scattering inside the subbands (dashed arrows) as well as between the subbands (dotted arrow).
	}
	\label{fig2}
\end{figure}

The Hamiltonian is conveniently presented as
\begin{equation}
\mathcal H+\delta \mathcal H + \mathcal H_{\rm Z}=\varepsilon_k + 
\bm \sigma\cdot \bm B^{\rm eff}(\bm k),
\end{equation}
where the spin-independent  term reads $ \varepsilon_k={\hbar^2 k^2/(2m)}$, and the effective magnetic field reads 
\begin{equation}
\bm B^{\rm eff}(\bm k)
= [\beta (k_x^2-k_y^2),2\alpha k_x k_y,\Delta].
\end{equation}
The electron energy $\varepsilon_{s\bm k}=\varepsilon_k + sB^{\rm eff}(\bm k)
$, where $s=\pm$ enumerates two spin subbands, and the effective magnetic field has a magnitude
$B^{\rm eff}=\abs{\bm B^{\rm eff}}$.
The isoenergetic contours plotted in Fig.~\ref{fig2} show that the spin degeneracy is lifted at finite $\alpha$ for all directions of $\bm k$ even at $\Delta=0$, which is confirmed also in first-principle calculations~\cite{Smejkal2020}.

The expectation values of the spin $x$ component in the eigenstates $\ket{s}$ are equal to $(s/2)n_{{\bm k},x}$, where ${\bm n_{\bm k} = \bm B^{\rm eff}(\bm k)/B^{\rm eff}(\bm k)}$. Therefore the electron spin orientation degree is calculated as follows
\begin{equation}
\label{sx_def}
s_x = {1\over N}\sum_{\bm k} \sum_{s=\pm}{s\over 2}n_{{\bm k},x} f_{s\bm k}.
\end{equation}
Here $f_{s\bm k}$ is the distribution function which satisfies the kinetic equation
\begin{equation}
\label{kin_eq}
{e\over \hbar}\bm E \cdot 
\bm \nabla_{\bm k} f_{s\bm k} = {2\pi\over \hbar}\sum_{\bm k',s'}\abs{V_{s'\bm k',s\bm k}}^2 \delta(\varepsilon_{s\bm k}-\varepsilon_{s'\bm k'}) (f_{s'\bm k'}-f_{s\bm k}).
\end{equation}
We consider elastic scattering by short-range randomly distributed defects $V(\bm r)=V_0\sum_i\delta(\bm r - \bm R_i)$, where $\bm R_i$ are the positions of the scatterers. In this case the scattering matrix element squared averaged over $\bm R_i$ reads
$\abs{V_{s'\bm k',s\bm k}}^2 = N_i V_0^2 \qty(1 + ss' \bm n_{\bm k}\cdot \bm n_{\bm k'})/2$,
where $N_i$ is a concentration of scatterers.
The kinetic Eq.~\eqref{kin_eq} accounts for scattering both inside the subbands ($s'=s$) and between them ($s'=-s$), Fig.~\ref{fig2}.

We search the distribution functions in the subbands iterating the kinetic equation in the electric field magnitude. As a result, we obtain a Taylor series expansion $f_{s\bm k}= f_0(\varepsilon_{s\bm k}) + \delta f_{s\bm k}+ \delta \tilde{f}_{s\bm k}$, where $f_0$ is the Fermi-Dirac distribution, $\delta f_{s\bm k} \propto E$, and $\delta \tilde{f}_{s\bm k} \propto E^2$.
%
%
In the absence of the spin-orbit splitting, the solution of this equation $\delta \tilde{f}_{\bm k}$ is given by Eq.~\eqref{align}, its angular dependence is shown  in Fig.~\ref{fig1} for $\bm E \parallel x$ and $\bm E \parallel y$.
The spin-orbit coupling makes the distribution different in the subbands, i.e. dependent on $s$.
For $\bm E \parallel x$, 
we have
\begin{align}
\label{kin_eq_1}
&k_{{\rm dr},x}^2  \partial^2_{k_x^2}f_0(\varepsilon_{s\bm k}) = \delta \tilde{f}_{s\bm k} \nonumber
\\
&-{1\over g} \sum_{\bm k',s'}{1 + ss' \bm n_{\bm k}\cdot \bm n_{\bm k'}\over 2}\delta(\varepsilon_{s\bm k}-\varepsilon_{s'\bm k'}) \delta \tilde{f}_{s'\bm k'}, 
\end{align}
where $g=m/(2\pi\hbar^2)$ is the density of states per spin in the absence of the spin splitting, $\bm k_{\rm dr}=e\bm E\tau/\hbar$ is the drift wavevector, and 
$1/\tau=(2\pi/\hbar)gN_i V_0^2$ is the relaxation rate.
We introduce the spin density at a given energy: 
\begin{equation}
\label{Sx_sx}
S_x(k) = \sum_{s=\pm}{s\over 2}\expval{n_{{\bm k},x} \delta f_{s\bm k}},
\quad
s_x = {1\over N}\sum_{\bm k} S_x(k),
\end{equation}
where angular brackets mean averaging over the angle $\varphi_{\bm k}$.
Then the equation for $S_x(k)$ is obtained from the kinetic Eq.~\eqref{kin_eq_1} by multiplying it by $(s/2)n_{{\bm k},x}$, summing over $s$ and averaging by directions of $\bm k$:
\begin{align}
&k_{{\rm dr},x}^2  \sum_{s=\pm}{s\over 2}\expval{n_{{\bm k},x} \partial^2_{k_x^2}f_0(\varepsilon_{s\bm k})} = S_x(k) 
\\ &-
\sum_{s=\pm}{s\over 2g}\expval{n_{{\bm k},x} \sum_{\bm k',s'}{1 + ss' \bm n_{\bm k}\cdot \bm n_{\bm k'}\over 2}\delta(\varepsilon_{s\bm k}-\varepsilon_{s'\bm k'})} \delta \tilde{f}_{s'\bm k'}. \nonumber
\end{align}
The term with `1' cancels after averaging over $\varphi_{\bm k}$, and then  $\delta(\varepsilon_{s\bm k}-\varepsilon_{s'\bm k'})$ is substituted by $\delta(\varepsilon_k-\varepsilon_{s'\bm k'})$ owing to summation over $s$. This yields
\begin{align}
k_{{\rm dr},x}^2 & \expval{n_{{\bm k},x} \partial^2_{k_x^2}f_0'(\varepsilon_k)B^{\rm eff}(\bm k)} = S_x(k) 
\\ &-{1\over g} \expval{n_{{\bm k},x}^2} \sum_{\bm k',s'}{ s' n_{\bm k',x}\over 2}\delta(\varepsilon_k-\varepsilon_{s'\bm k'}) \delta \tilde{f}_{s'\bm k'}. \nonumber
\end{align}
Now using 
$$\delta(\varepsilon_k-\varepsilon_{s'\bm k'})=\delta(\varepsilon_k-\varepsilon_{k'})+s'B^{\rm eff}(\bm k')\partial_{\varepsilon_k}\delta(\varepsilon_k-\varepsilon_{k'}),
$$ we  get
\begin{align}
&  k_{{\rm dr},x}^2 \expval{n_{{\bm k},x} \partial^2_{k_x^2}f_0'(\varepsilon_k)B^{\rm eff}(\bm k)} = S_x(k) \qty(1-\expval{n_{{\bm k},x}^2}) 
\\ &-{1\over g} \expval{n_{{\bm k},x}^2} \partial_{\varepsilon_k}\sum_{\bm k'}\delta(\varepsilon_k-\varepsilon_{k'})n_{\bm k',x} B^{\rm eff}(\bm k') {1\over 2}\sum_{s'}\delta \tilde{f}_{s'\bm k'}.\nonumber
\end{align}
The last sum is a spin-independent part of the distribution function which can be taken in the form
\begin{equation}
{1\over 2}\sum_{s'}\delta \tilde{f}_{s'\bm k'} = k_{{\rm dr},x}^2\partial^2_{k_x'^2}f_0(\varepsilon_{k'}),
\end{equation}
because only its anisotropic part contributes. As a result, we have
\begin{align}
S_x(k) \qty(1-\expval{n_{{\bm k},x}^2}) =& k_{{\rm dr},x}^2  \biggl[ \expval{n_{{\bm k},x} \partial^2_{k_x^2}f_0'(\varepsilon_k)B^{\rm eff}(\bm k)}
\\&+ \expval{n_{{\bm k},x}^2} \partial_{\varepsilon_k} \expval{B_x^{\rm eff}(\bm k)\partial^2_{k_x^2}f_0(\varepsilon_{k}) } \biggr]. \nonumber
\end{align}
Summation over $\bm k$, see Eq.~\eqref{Sx_sx}, yields for the spin orientation degree
\begin{align}
s_x = {k_{{\rm dr},x}^2\over N}\sum_{\bm k}  & \Biggl[ {n_{{\bm k},x} \over 1-\expval{n_{{\bm k},x}^2}} \partial^2_{k_x^2}f_0'(\varepsilon_k)B^{\rm eff}(\bm k)
\\& + {\expval{n_{{\bm k},x}^2} \over 1-\expval{n_{{\bm k},x}^2}} \partial_{\varepsilon_k} f_0''(\varepsilon_k)\expval{B^{\rm eff}_x(\bm k)(\hbar v_x)^2  }\Biggr].\nonumber
\end{align}
Since $\expval{B^{\rm eff}_x(\bm k)(\hbar v_x)^2  }=\beta\varepsilon_k^2$, integration by parts
 gives (for an arbitrary orientation of the electric field)
%
%
%
%
%
%
\begin{align}
\label{s_fin}
s_x = -{k_{{\rm dr},x}^2-k_{{\rm dr},y}^2 \over 2\varepsilon_{\rm F}} &   \Biggl[\expval{B^{\rm eff}(\bm k)
\partial^2_{k_x^2} {n_{{\bm k},x} \over 1-\expval{n_{{\bm k},x}^2}} }_{\rm F} \nonumber
\\
+&\beta \partial_{\varepsilon_{\rm F}} \qty( \varepsilon_{\rm F}^2 \partial_{\varepsilon_{\rm F}}{\expval{n_{{\bm k},x}^2}_{\rm F} \over 1-\expval{n_{{\bm k},x}^2}}_{\rm F} )  \Biggr],
\end{align}
where $\expval{\ldots}_{\rm F}$ means averaging over the Fermi circle.
%

The obtained expression Eq.~\eqref{s_fin} gives the electron spin polarization at arbitrary relation between $\beta$, $\alpha$ and $\Delta$.

\mysection{Discussion}
First, we consider the electrical spin orientation in the absence of magnetization, when $\Delta=0$. In this case $\expval{n_{{\bm k},x}^2}=\beta/(\alpha+\beta)$ is independent of the energy $\varepsilon_k$, therefore the second term in Eq.~\eqref{s_fin} equals to zero, and we get
\begin{equation}
\label{s_zero_Delta}
s_x(\Delta=0)=  -{(\beta-\alpha) (k_{{\rm dr},x}^2-k_{{\rm dr},y}^2) \over 2\varepsilon_{\rm F}}.
\end{equation}
Interestingly, at isotropic spin splitting, when $\alpha=\beta$, the spin orientation disappears.
In the limit $\alpha \to 0$ 
we obtain $s_x = Q (j_x^2 - j_y^2)$ in accordance with phenomenological Eq.~\eqref{s_j_phenom}, where 
\begin{equation}
 Q =-{\beta \over 2\varepsilon_{\rm F}} \qty({m\over e\hbar N})^2.
\end{equation}
Comparing with the estimate Eq.~\eqref{Q_estimate} we see that it is correct up to the factor $-1/2$ found from the microscopic calculation.
The spin orientation degree scales with the current and electron densities as $s\propto j^2/N^3$. For $\beta=1$~eV\AA$^2$ and $m=m_0$ we obtain $s \approx 6.2\times 10^{-4}$ at $j=1{\rm mA/cm}$ and $N=10^{11}{\rm cm}^{-2}$.

\begin{figure}[t]
	\centering \includegraphics[width=\linewidth]{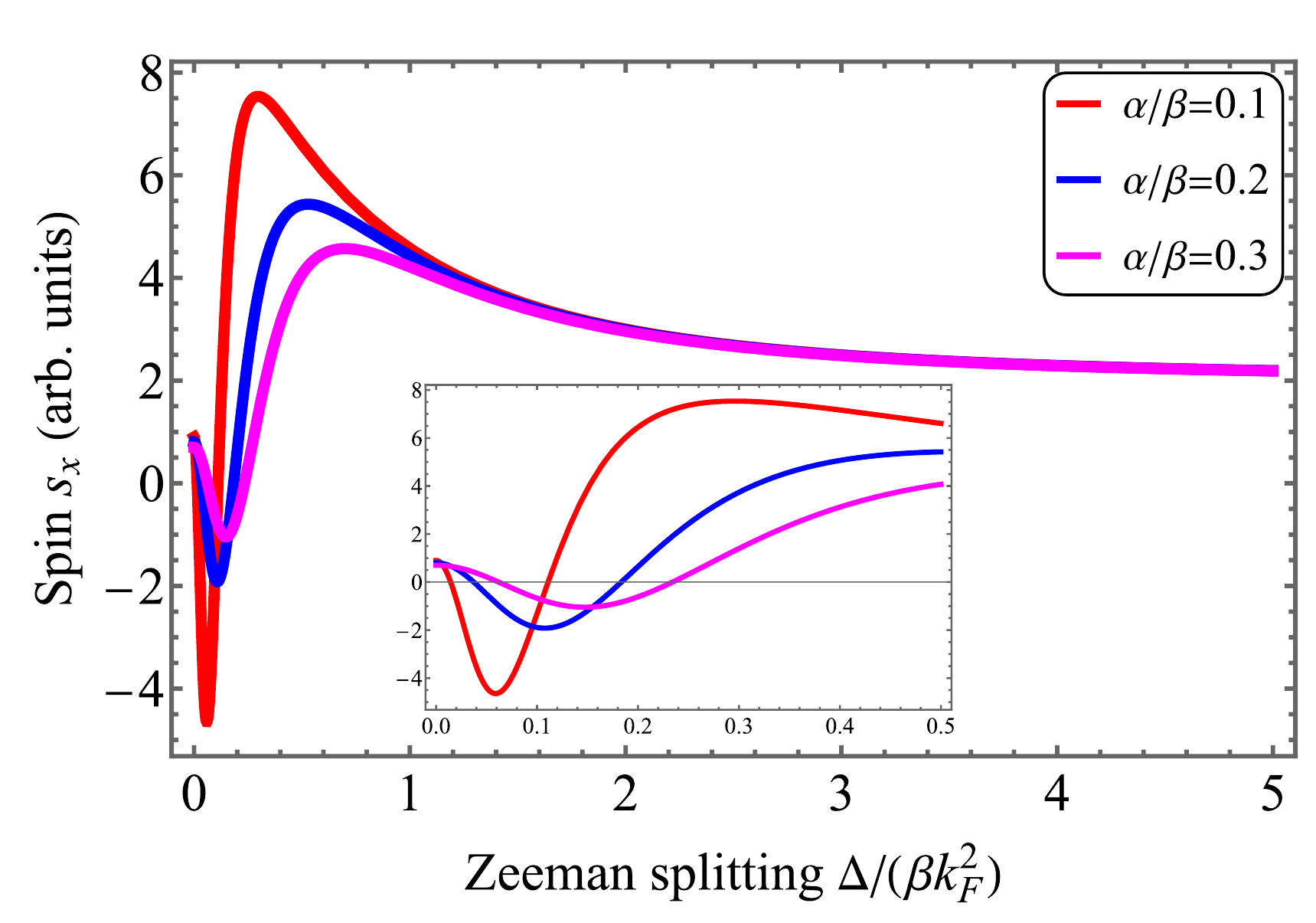} 
	\caption{ 
	Effect of Zeeman splitting on the electrical spin orientation. The electron spin $s_x$ is normalized to the value at $\Delta=0$, Eq.~\eqref{s_zero_Delta}. Inset shows the low-splitting range.
	}
	\label{Fig_Zeeman}
\end{figure}

Now we consider the effect of Zeeman splitting. In this case, we have
\[\expval{n_{{\bm k},x}^2} = {\beta^2k^4\over \beta^2k^4 + \Delta^2+\sqrt{(\beta^2k^4 + \Delta^2)(\alpha^2k^4 + \Delta^2)}},
\]
and angular averaging in the first term of Eq.~\eqref{s_fin} is performed numerically.
The dependence of the electron spin density on the Zeeman splitting is shown in Fig.~\ref{Fig_Zeeman}.

At low Zeeman splitting, the spin has sharp changes with a minimum at $\Delta \sim \alpha k_{\rm F}^2/2$, where $k_{\rm F}$ is the Fermi wavevector -- see inset to Fig~\ref{Fig_Zeeman}. With increasing $\Delta$, the spin approaches maximum with a value controlled by the ratio $\alpha/\beta$. At large Zeeman splitting, the spin tends to the limit $s_x(\Delta \to \infty)= -{(\beta-\alpha) k_{\rm dr}^2 /\varepsilon_{\rm F}}$. This value is twice larger than the spin orientation degree at $\Delta=0$, Eq.~\eqref{s_zero_Delta}.

The odd in the magnetization perpendicular component $s_y \propto E_x^2 \Delta$ is also allowed by symmetry. However, it appears in the next order in ${\hbar /(B^{\rm eff} \tau)\ll 1}$~\cite{Golub2025}, therefore it is absent in the spin-incoherent approach of well-separated spin subbands, Fig.~\ref{fig2}.

A similar consideration shows that the electron spin orientation by electric current is also possible in $g$- and $i$-wave altermagnets. The spin orientation degrees have higher-order nonlinearity: $s \propto j^4$ and $s \propto j^6$ and  vary under rotation of the electric current as 4th and 6th angular harmonic for $g$- and $i$-wave symmetries, respectively.

{We point out that the spin-quadratic-momentum locking, such as that in Eq.~\eqref{H}, does not have to emerge only from nonrelativistic $d$-wave altermagnetism (e.g. rutile~\cite{Smejkal2020} or Lieb lattice systems~\cite{Smejkal2022,Mazin2023,Fernandes2024}) but can materialize also from a combined effect of relativistic spin-orbit coupling and nonrelativistic $g$-wave exchange forces in altermagnets~\cite{Krempasky2024,Fernandes2024,Belashchenko2025}. In fact,
$\mathcal H + \delta\mathcal H =  \beta(k_x^2-k_y^2)\sigma_{x} + 2\alpha k_x k_y \sigma_{y}
$ (with $\alpha=\beta$) describes lowest order $\bm k\cdot \bm p$ model of $k_{z}=0$ plane of $B_{2g}^{-}$ altermagnetic representation of $D_{6}$ hexagonal group; e.g. magnetic point group $6'/m'mm'$~\cite{Fernandes2024}. This Hamiltonian is thus relevant for altermagnets with NiAs structure such as MnTe~\cite{Smejkal2022a,Krempasky2024} and CrSb~\cite{Smejkal2022a,Reimers2024} with Néel vector set along c-axis.}

\mysection{Summary}
A theory of electron spin orientation by electric current in altermagnets is developed. The electron spin in $d$-wave altermagnets is quadratic in the electric current and has a characteristic angular dependence with variation of the current direction. In the approximation of well-resolved spin subbands, kinetic theory is used to calculate the electron spin, taking into account both inter- and intra-subband elastic scattering. It is shown that perpendicular magnetization leads to a multiple enhancement of the electron spin.

\mysection{Acknowledgments}
L.~E.~G. is supported by the Deutsche
Forschungsgemeinschaft (DFG, German Research
Foundation) Project No. Ga501/18. 
L.~Š. acknowledges support from the ERC Starting Grant No. 101165122.

\bibliography{CISP_altermagn.bib}

\end{document}